# X-ray, UV and Optical Observations of Classical Cepheids: New Insights into Cepheid Evolution, and the Heating and Dynamics of Their Atmospheres

Scott G. Engle[1†] & Edward F. Guinan[1]
[1]Department of Astronomy and Astrophysics, Villanova University, Villanova, PA 19085, USA

[†]Corresponding Author
E-mail: scott.engle@villanova.edu
Tel: 610-519-4820   Fax: +

**ABSTRACT**

To broaden the understanding of Classical Cepheid structure, evolution and atmospheres, we have extended our continuing *Secret lives of Cepheids* (*SLiC*) program by obtaining *XMM/Chandra* X-ray observations, and *HST/COS* (Cosmic Origins Spectrograph) FUV-UV spectra of the bright, nearby Cepheids Polaris, δ Cep and β Dor. Previous studies made with the *International Ultraviolet Explorer (IUE)* showed a limited number of UV emission lines in Cepheids. The well-known problem presented by scattered light contamination in *IUE* spectra for bright stars, along with the excellent sensitivity & resolution combination offered by *HST/COS*, motivated this study, and the spectra obtained were much more rich and complex than we had ever anticipated. Numerous emission lines, indicating $10^4$ K up to ~$3\times10^5$ K plasmas, have been observed, showing Cepheids to have complex, dynamic outer atmospheres that also vary with the photospheric pulsation period. The FUV line emissions peak in the phase range $\varphi \approx 0.8$–$1.0$ and vary by factors as large as 10×. A more complete picture of Cepheid outer atmospheres is accomplished when the *HST/COS* results are combined with X-ray observations that we have obtained of the same stars with *XMM-Newton* & *Chandra*. The Cepheids detected to date have X-ray luminosities of log $L_X \approx 28.5$–$29.1$ ergs/sec, and plasma temperatures in the $2$–$8\times10^6$ K range. Given the phase-timing of the enhanced emissions, the most plausible explanation is the formation of a pulsation-induced shocks that excite (and heat) the atmospheric plasmas surrounding the photosphere. A pulsation-driven $\alpha^2$ equivalent dynamo mechanism (Chabrier & Küker 2006) is also a viable and interesting alternative. However, the tight phase-space of enhanced emission (peaking near $0.8$–$1.0\varphi$) favor the shock heating mechanism hypothesis.

## 1. A BRIEF INTRODUCTION TO CEPHEIDS

*Classical Cepheids* (in this paper shall be hereafter referred to simply as *Cepheids*) are a fundamentally important class of pulsating variable stars. Cepheids are moderate-mass (typically 4–10 $M_\odot$), luminous (~$10^3$–$10^5$ $L_\odot$), white - yellow (spectral types of approximately F6–K2) supergiants (luminosity classes of Ia, Ib and II) whose radial pulsations produce periodic variations in radius, temperature and, consequently, brightness. The pulsation periods of Cepheids range from just under 2-days to as long as ~45-days (for SV Vul, but as long as ~60-days depending on the classification of S Vul). The *dominant* (but not the only) pulsation mechanism at work in Cepheids is the *kappa (κ) mechanism* (so named for κ – the coefficient of absorption/opacity). The κ mechanism operates within the Cepheids, specifically within partially ionized He II zones in the upper layers of Cepheids. This ionization mechanism causes the zone to absorb energy during compression and then release it during expansion. This behavior is sometimes referred to as the *Eddington* *valve*, since he was the first to propose such a theory, although he proposed hydrogen as the responsible element (Eddington 1941). For Cepheid pulsations to be successfully driven, two basic criteria must be satisfied: a sufficient amount of ionizing material (helium, in the case of Cepheids) must exist, and at a layer within the star where a transition is occurring from adiabatic to non-adiabatic behavior (Cox 1985). Specifically, a star must contain a concentration of at least 10–15% helium, roughly half of which must be ionized (Kukarkin 1975). Also, the second ionization of helium (He II to He III) begins at 35000–55000K (Bohm-Vitense 1958). The stellar depth at which these temperatures exist progresses from nearer to the surface in hotter (earlier-type) stars down to depths nearer the core in cooler (later-type) stars. Because of the stringent requirements on ionization zone depth, combined with the dampening effect of strong convection in cooler stars, Cepheid pulsations can only be maintained in a specific range of effective temperature. This range is defined as the steep-diagonal "classical instability strip" where Cepheids are found on the HR diagram (see Fig. 1).

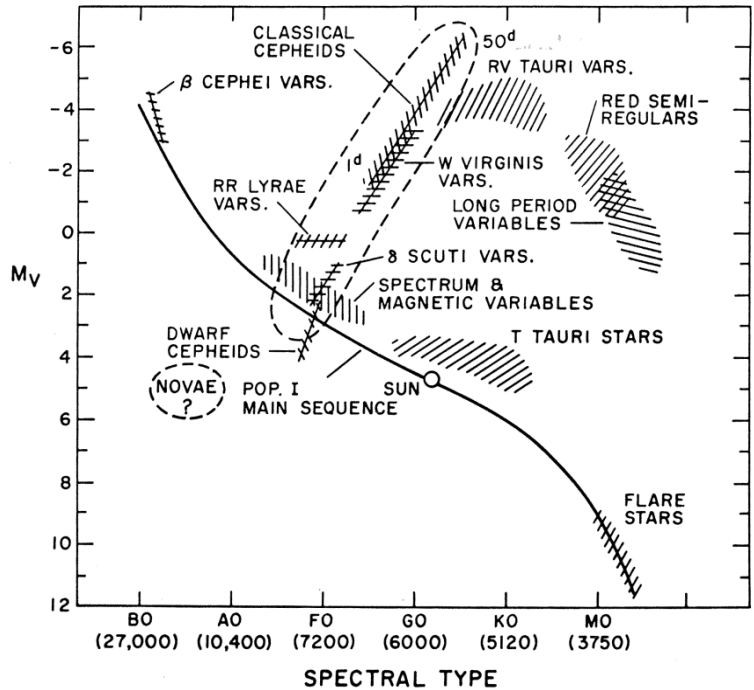

**Fig. 1** – Hertzsprung-Russel diagram showing the locations of various types of intrinsic variables, including Classical Cepheids at the top-center of the diagram (Cox 1974).

The Astronomical importance of Cepheids is long-standing. Since the discovery of the Period-Luminosity Relationship (The "Leavitt Law") over a century ago (Leavitt 1908), Cepheids have become not just a cornerstone of the Extragalactic Distance Scale (see – Freedman & Madore 2010, but also are excellent laboratories to study radial pulsations, stellar structure and evolution of these astrophysically and cosmologically important stars. The scope of Cepheid studies has shifted somewhat over the years, specifically away from the brightest Cepheids, of which CCD observations are very difficult. As discussed below, however, the brightest Cepheids are nonetheless very valuable because they possess long timelines of observational studies. Combining these archival data with continuing studies permits the possible delineation of stellar evolution along human timescales – e.g. changes in period, amplitude and possibly even average brightness. Additionally, due to the proximity of the brightest Cepheids, these stars are the most suitable targets for new and advanced instruments, as we will illustrate in the UV and X-ray regimes. In summary: to obtain a true understanding of their atmospheric structure and activity, the continuation and expansion of the observational database is necessary.

## 2. THE SECRET LIVES OF CEPHEIDS

The *Secret Lives of Cepheids* (*SLiC*) program is a comprehensive study of Cepheid behavior, evolution, pulsations, atmospheres, heating, dynamics, shocks and winds that we have been carrying out for several years. This program currently includes ~15 of the brighter Cepheids, covering a representative range of pulsation properties. The program spans almost the entire electromagnetic spectrum, from X-ray observations (*XMM-Newton & Chandra*) and FUV/UV spectra (*Hubble Space Telescope (HST), the International Ultraviolet Explorer (IUE) and Far Ultraviolet Spectroscopic Explorer (FUSE)*, to the ground-based photoelectric photometry we continue to gather (along with CCD photometry of some fainter Cepheids), to IR–Far-IR spectra (*Spitzer Space Telescope*). Taken together, these data have revealed Cepheids to be increasingly complex, with surprising levels of activity and variability. At optical wavelengths, we continue to monitor Cepheid light curves for evidence of evolutionary changes in morphology, average brightness and times of maximum light. For example, in several Cepheids, the pulsation period is known to change (increase or decrease) significantly, by as much as 200+ sec/year (Turner & Berdnikov 2004). The low light-amplitude Cepheid Polaris – a program star – has been undergoing a +4.4 sec/yr increase in its pulsation period (Neilson et al. 2012). Additionally, Polaris' light amplitude decreased for nearly a century, and it almost ceased to be a Cepheid in the early 1990s when V-band light amplitude reached minimum value of ~0.02-mag (see Kamper & Fernie 1998). Our photometry first indicated the recovery of Polaris' light amplitude in the early 2000s (Engle, Guinan & Koch 2004) and subsequent studies have confirmed our initial findings (Spreckley & Stevens 2008). Our 2011/12 photometry indicates the light amplitude continues its increase, with a V-band amplitude of ~0.062-mag (see Fig. 2). Also, analysis of historic photometry and magnitude estimates of Polaris back to the *Almagest* (Ptolemy: circa ~130 CE; Sufi: circa ~960 CE – both listing Polaris as a "3$^{rd}$" mag star, while today Polaris is near V ≈ 2.0 mag) indicates that its luminosity could have increased by nearly a magnitude (Engle, Guinan & Koch 2004) over this time. However, a great deal work is still required to reliably calibrate the historic data to the modern V-band. Finally, O-C data for Polaris show its period to currently be increasing at a rate of ~4.47±1.25 sec/year, which could be indicative of enhanced mass loss from the Cepheid (Neilson et al. 2012). Another, possible more extreme example of real-time Cepheid evolution is the case of Hubble's V19 in M33, which was discovered as a 54.7-day full-amplitude (~1.0 mag in blue) Cepheid, but has since brightened by ~0.5-mag and apparently stopped pulsating (Macri et al. 2001, Engle et al. 2011).

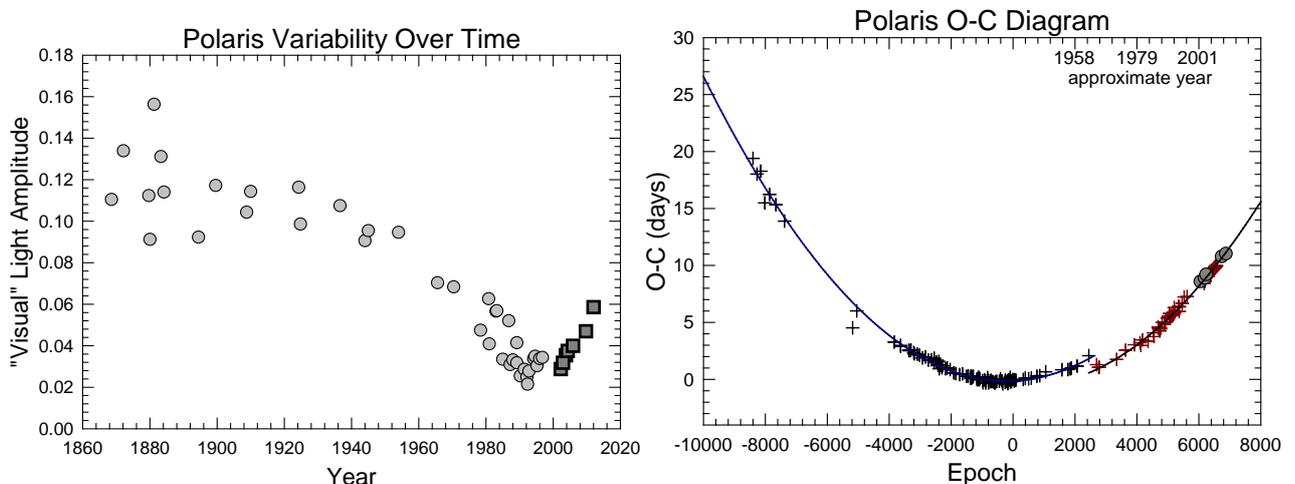

**Fig. 2 – Above –** A plot of the V-band (or corrected visual magnitude) light amplitude of Polaris is shown, going back to the late 1800s. The decline in light amplitude beginning in the mid 1900s is clearly visible, as is the resurgence of the light amplitude during the 21$^{st}$ century. The gray dots are archival data (taken mostly from Kamper & Fernie 1998 and Arellano Ferro 1983), and the dark gray squares are data from our program.
**Fig. 3 – Above Right & Right –** O-C data for Polaris going back to 1844/5 is shown. Crosses represent data from Turner et al. 2005 and Spreckley & Stevens 2008, while the gray dots represent data from our program. The data has been broken into two sets (pre-1963 and post-1965), consistent with an apparent rapid change in period during 1963–1966 (Epoch 2400–2600). Weighted, least squares quadratic fits have been run to the data sets, returning a period increase of 4.44±0.19 sec/year for the early data and 4.47±1.25 sec/year for the later data.

In addition to the evolutionary studies, we have begun investigating the atmospheric dynamics and heating of Cepheids using X-ray–UV data (the focus of this paper). Specifically, these data reveal plasmas in Cepheid atmospheres with temperatures of $10^3$–$10^6$ K (Engle et al. 2009). Also, the hot plasmas are variable in accord with the pulsations of the Cepheids. The very long exposures required for good X-ray characterization have slowed progress in that wavelength range, but due to the excellent *HST/COS* sensitivity, the UV–FUV situation is rather different.

### 3. PREVIOUS UV AND FUV STUDIES

The history of Cepheid "super-Photospheric" (plasmas with a temperature above that of the photosphere) studies is sparse. The earliest of such studies appears to be that of Kraft (1957), who studied the Ca II *H* (3968.5 Å) and *K* (3933.7Å) lines in a large number (20+) of brighter Cepheids. The Ca II H&K lines originate in plasmas with temperatures in the low $10^4$ K range; similar to a low temperature chromosphere in solar-type stars. Kraft noted that the Ca II emissions peaked in Cepheids around $\varphi \approx 0.8$–$1.0$ (just after the Cepheid has begun to expand from minimum radius). Due to the star's expansion, a shock is expected to pass through the Cepheid photosphere at this phase, which is sometimes referred to as the "piston phase" of Cepheids. From this, Kraft concluded that "the transitory development of Ca II *H&K* emission in Classical Cepheids is associated with the appearance of hot material low in the atmosphere. These hot gases are invariably linked with the onset of a new impulse." This important study laid the groundwork for Cepheid atmospheric studies, but the Ca II lines can only probe cooler atmospheric plasmas. More than two decades would pass before an investigation of higher temperature plasmas around Cepheids was conducted.

Schmidt & Parsons (1982; 1984a; 1984b) made perhaps the most thorough and revealing study of Cepheid atmospheres using *IUE* spectra. The wavelength range of *IUE* (~1200–3200Å) covers several important emission lines with temperatures of $10^4$–$10^5$ K, equivalent to those found in solar-type chromospheres and transition regions. In accord with the results of Kraft (1957), Schmidt & Parsons found these emissions to be variable, peaking shortly before maximum light (which is also shortly after minimum radius). At this phase, the stellar photosphere begins to expand again and, as such, this is referred to as the *piston phase* of a Cepheid. Transition region lines were also found, but were not as strong as chromospheric emissions and were more easily contaminated by the photospheric continuum in all but the longest exposed (and ideally phase-space located) spectra. Still, the sporadic detections of such lines served as evidence that Cepheid atmospheres could be more complex (and possibly hotter), and that perhaps a more modern, and higher resolution, instrument could reveal such lines in a more concrete way.

Motivated by these results, an analysis of *FUSE* (~920–1190 Å) spectra was carried out, and new observations were successfully proposed for. Sadly, although multiple *FUSE* observations of additional Cepheids were approved, just one observation (for β Dor) was carried out before the fatal failure of the mission's guidance system. The "*FUSE* Cepheid database" therefore includes only Polaris and β Dor (and only β Dor has multiple observations, but of somewhat poor phase coverage). However, both the C III 977Å and O VI 1032/1038Å line emissions were not only found in the spectra of both Cepheids, but were found to increase in strength at the piston phase of β Dor ($\varphi \approx 0.8$, Engle et al. 2009). This result is in agreement with the phase of peak N V 1240 Å emission (N V forming at comparable temperatures to O VI), found from our *HST/COS* observations (discussed in the following section). The C III emission and variability was somewhat expected, given that it forms at similar temperatures (~4–5×$10^4$ K) to emission lines studied in the archival *IUE* spectra. However, the discovery of the O VI lines was a pleasant surprise, given that they form in plasmas with temperatures of ~3–4×$10^5$ K (Redfield et al. 2002). The O VI lines were, at that time, the hottest plasma emissions observed from a Cepheid. This study raised the questions: just how hot is a Cepheid's atmosphere, and how is it being heated? Given the phase-timing of the enhanced emissions, the most plausible explanation is the formation of a shock that excites the atmospheric plasmas surrounding the photosphere. A pulsation-driven $\alpha^2$ equivalent dynamo mechanism (Chabrier & Küker 2006) is also a viable and interesting alternative. However, the tight phase-space of enhanced FUV emissions lends found

## 4. HST/COS FUV SPECTROPHOTOMETRY: NEW INSIGHTS IN THE STRUCTURE, DYNAMICS & HEATING OF CEPHEID ATMOSPHERES

During 2009-10, we were awarded 20 *HST* orbits for *COS* spectroscopy of three Cepheids: 2 visits (2 orbits/visit) for Polaris, and 16 visits (1 orbit/visit) for δ Cep & β Dor. Because of the scattered light contamination (Fig. 4), *IUE* spectra of these stars only unambiguously show the only the strongest emission lines (if any). In the case of Polaris, as is shown in the Fig. 4, there was not even certainty that emission lines were present in the FUV spectra; these were only possible lines present in the *IUE* data. *HST/COS* spectra of Cepheids, however, display a wealth of emission lines not detected in archival the *IUE* spectra (Fig. 2). These lines define rich and complex Cepheid atmospheres and, as is well known, offer excellent atmospheric diagnostic potential (Linsky et al. 1995 and references therein), since different line species originate in plasmas of specific temperatures. Also, emission line strengths and ratios, as well as line broadening and radial velocities, when measured over the stars' pulsation cycles, offer important atmospheric diagnostics. FUV emission line strengths and ratios (along changes in radial velocity with pulsation) can also help distinguish Cepheid supergiant atmospheric emissions from those of possible unresolved main sequence companions (if present).

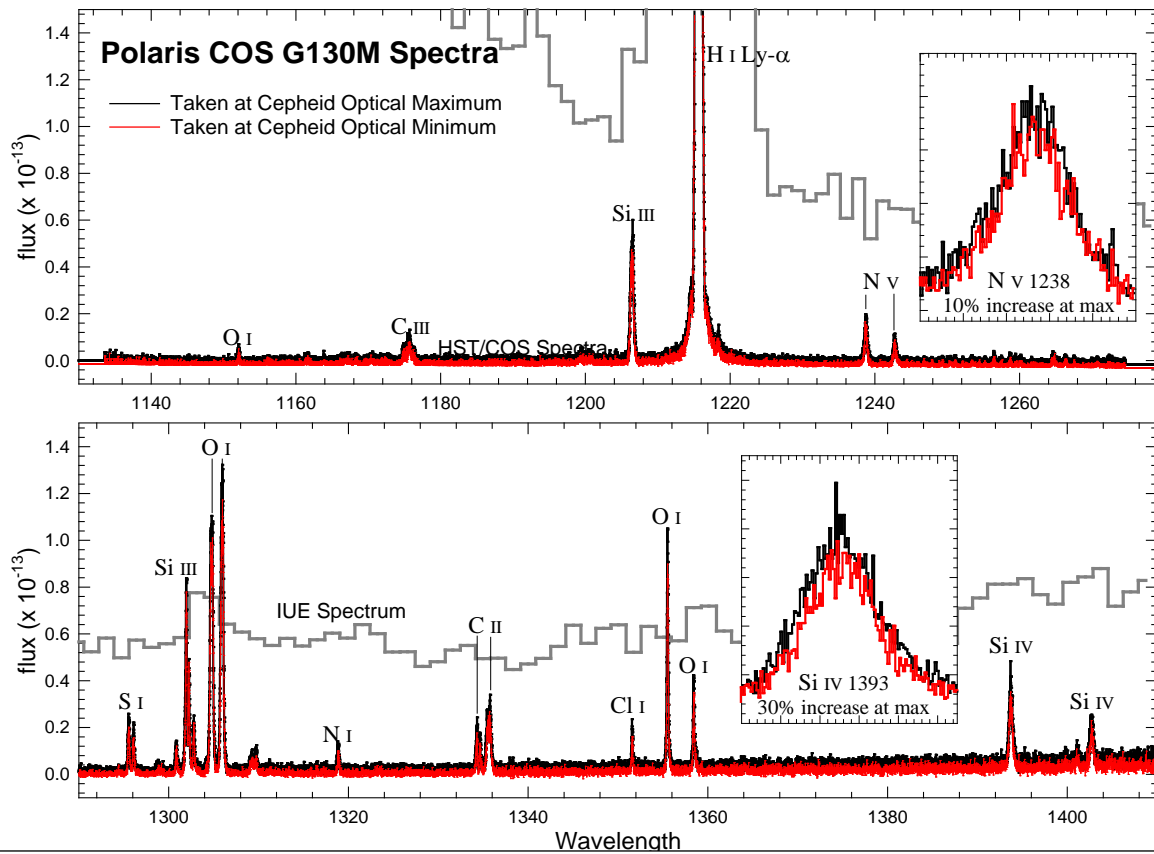

**Fig. 4 – Above –** The two *HST/COS* G130M spectra of Polaris are shown. The black spectrum was taken near the phase of maximum light, and the red spectrum was taken near minimum light. Many of the important emission features are labeled. The thick gray line above the COS spectra is the best-exposed IUE spectrum of Polaris. As shown, it is heavily affected by scattered light, especially below 1180Å where it rises above the scale of the plot.

As an illustration of the initial results of the FUV program, the nearly complete *COS* light curves of δ Cep are shown in Fig. 5. As shown in these plots, a few features are striking: the specific phases at which line emissions begin to increase, the abruptness of the increase, and also the apparent phase-difference between various emission lines (notably in the most energetic emission feature observed: N V). In comparison to the included radial velocity (RV) plots for each Cepheid, we see good correlation between the phase where emissions begin to peak and the beginning of the "piston phase" where the Cepheid photosphere begins to expand again. Because of this, we now realize that we most likely missed the expected phase of maximum FUV line emission in Polaris. One interesting result shown from our phase coverage of δ Cep (see Fig. 5) is that the lower temperature plasma emission lines such as O I 1358Å, appear to peak first (Phase ≈ 0.86φ) while the hotter Si IV 1400Å and N V 1240Å emission lines reach maximum strength near ~0.94φ (interpolated) and ~0.0φ, respectively. But, as discussed by Bohm-Vitense & Love (1994), the emissions from the hottest plasmas are expected to peak first in the case of shock-heating, followed by the line emissions from the cooling plasmas in the post-shock regions.

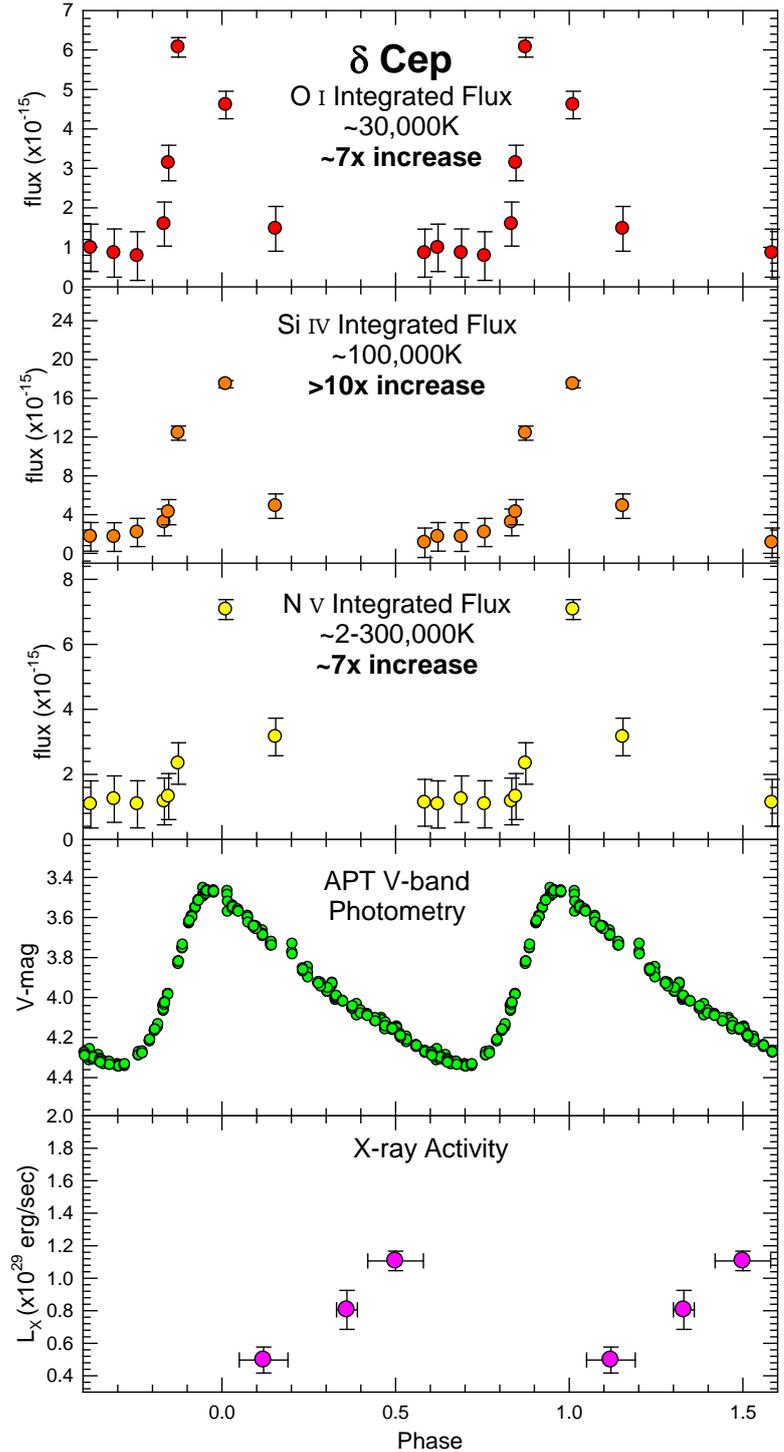

**Fig. 5 – Above Right –** Integrated fluxes of 3 important emission lines observed in δ Cep with *COS* - top three panels. The fourth panel (green points) is the V-band light curve from our continuing photometry. The bottom panel plots the X-ray luminosities for δ Cep to date. The UV lines trace out $10^5$ K plasma activity, while the X-ray data traces the crucial $10^6$ K plasmas. One can easily see the tight phase-space of enhanced UV emission. Although X-ray variability can also be seen, the curve is clearly incomplete. We hope to fill this curve in with future data.

# 5. THE (ALMOST) "UNDISCOVERED COUNTRY" OF CEPHEID X-RAY ACTIVITY

UV line emissions from $10^6$K (MK) plasmas are rare, typically weak, and only appear in the most active of stars (e.g. see Linsky et al. 1998 for a discussion of the coronal *Fe XXI* 1354Å emission line for Capella). Thus, to detect and study Cepheids at these high temperatures, X-ray observations are needed. Even though X-rays were predicted for Cepheids (from pulsation-induced shocks) in the mid-1990s (Sasselov & Lester 1994), the failure of previous efforts to detect X-rays with pointed Einstein and ROSAT, reinforced the idea that Cepheids are not significant X-ray sources. Even with $log$ $L_X \approx 29$ erg/s, the problem with detecting X-rays from Cepheids is that these stars (except for Polaris at ~133 pc – van Leeuwen 2007) are far away at d > 250 pc. Thus, they are expected to have relatively weak X-ray fluxes ($f_X < 10^{-14}$ ergs/s/cm$^2$). Though Polaris was detected on the 3σ level in a *ROSAT/HRI* archival image (several years after the observation was carried out – see Evans et al. 2007), definitive detection of X-rays from Polaris and other "nearby" Cepheids had to wait for the arrival of powerful X-ray observatories like *XMM-Newton* and *Chandra*.

Accordingly, we successfully obtained *Chandra* (PI: Evans) and *XMM-Newton* (PI: Guinan) observations of multiple Cepheids – Polaris, δ Cep, β Dor, SU Cas and *ℓ* Car have been observed so far. The nearest three Cepheids – Polaris, δ Cep & β Dor were detected in X-rays (Fig. 6: $log$ $L_X \approx 28.6–29.0$ ergs/s). The short period Cepheid SU Cas (P = 1.95-d; d = 421±36 pc: Turner et al. 2012) and the long period Cepheid *ℓ* Car (P = 35.5-d; d = 498±55 pc: Benedict et al. 2007) were not detected. Upper X-ray luminosity limits of log $L_X$ < 29.6 and 29.5 ergs/s were estimated for *ℓ* Car and SU Cas, respectively. This means that SU Cas and *ℓ* Car could indeed be X-ray sources but too distant to be detected with the <50-ksec *XMM-Newton* exposures. When we began these X-ray studies a few years ago, we often encountered the argument that unresolved companions were responsible for the activity we assign to Cepheids. And of course this is possible since Cepheids are young stars (~50–200 Myr) and main-sequence G-K-M companions (if present) would be coronal X-ray sources. However, our *HST/COS* results show plasmas up to $10^5$K that are variable with the Cepheids' pulsation periods – indicating these hot plasmas originate from the Cepheids. Also our (still sparse) X-ray observations of δ Cep and Polaris show variability possibly correlated with the stars' pulsation periods, as with the FUV emission lines. While we feel confident that Cepheids are also capable of producing the detected X-ray activity, we are awaiting further X-ray data to investigate the possible phase dependence of the X-ray emission on pulsation period. In addition to providing conclusive proof that Cepheids themselves are X-ray sources (and not the hypothetical young coronal companions previously mentioned), the additional X-ray observations would also shed more light on the formation and variability of hot (MK) Cepheid plasmas.

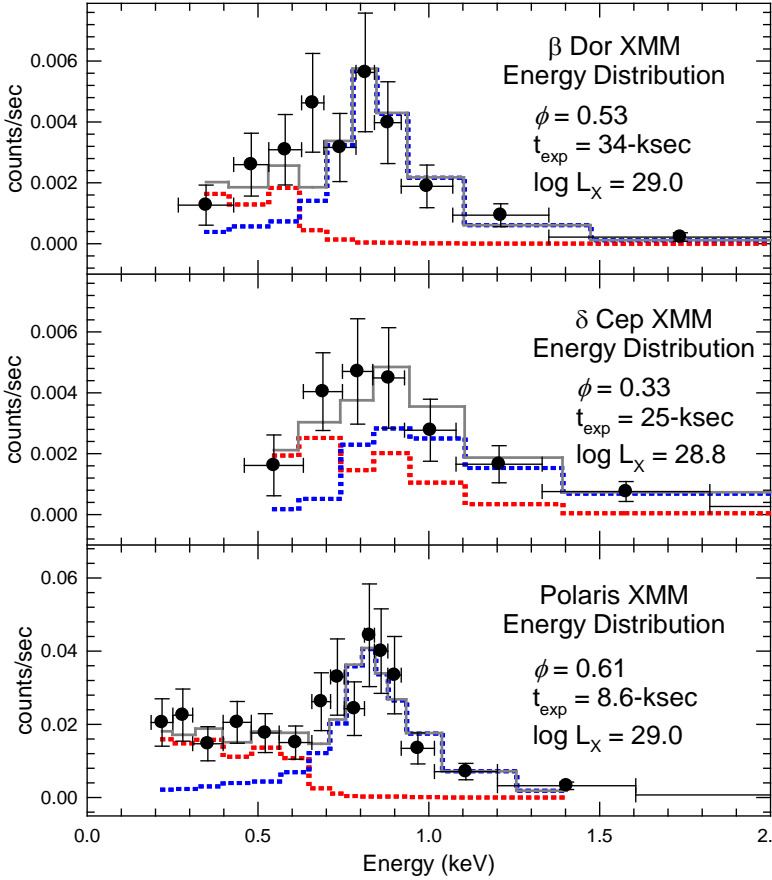

**Fig. 6 – Left** – X-ray energy distributions for the 3 Cepheids detected so far. Gray solid lines indicate the 2-T fit to the data, while blue and red dotted lines plot the individual contributions of the temperatures. We note that subsequent observations of these Cepheids have been of longer exposure times, and have started to reveal X-ray variability, as shown in Fig. 5.

At first glance, the mean $L_X$ values observed for Cepheids are on the order of 30–60× that of the mean solar value (adopting $log\ L_X \approx 1 \times 10^{29}$ ergs/s for Cepheids and $<L_X>_\odot \sim 2 \times 10^{27}$ ergs/s from DeWarf et al. 2010). However, the surface X-ray flux ($F_X$) and the $L_X/L_{bol}$ ratio, compared to the Sun, are much smaller ($log\ (L_X/L_{bol}) = -7.7$ [δ Cep] and $-6.3$ [Sun]). For example, in the case of δ Cep (adopting a radius of R = 44.5 $R_\odot$, L ~2000 $L_\odot$ (as given by Matthews et al. 2012), the surface X-ray flux is $F_X = 8.3 \times 10^2$ ergs/s/cm$^2$. The corresponding mean value for the Sun is $<F_X>_\odot \approx 4.1 \times 10^5$ ergs/s/cm$^2$. Thus, the average X-ray surface flux of the Sun is ~500× stronger than that of the Cepheid.

The most plausible (and very promising) mechanism for the variable, pulsation-phase dependent UV–FUV emissions, and observed X-ray emissions, is discussed by Sasselov & Lester (1994). They conclude Cepheids could have outer atmospheres heated by acoustic (or magnetic) shock wave dissipation, in addition to the transient heating by pulsation dynamics. From this theory (and observations of the He I 10830Å line), they predict average X-ray surface fluxes (0.05–1 keV) in the range of $F_X \approx 600$–11,000 ergs/cm$^2$/sec. Taking into account the stellar radii of Cepheids, this suggests average $L_X$ values of ~10$^{29}$ erg/s for Cepheids with P ≤ 10-days, in good agreement with what we have observed in Polaris, β Dor and δ Cep. This study was carried out in 1994, before X-rays were discovered from Cepheids (although pointed X-ray observations of the brighter (more nearby) Cepheids were carried out by *Einstein* and *ROSAT*). Our (so far) limited observations seem to confirm the theoretical expectations. Sasselov & Lester also predict the X-ray luminosities of Cepheids could vary by a factor of ~2.5× over the pulsation period. Indeed, as shown in Fig 5, the X-ray observations to date for δ Cep vary by (at least) by a factor of ~2. Additional phase coverage is needed to cover the expected X-ray maximum (during 0.7–1.0φ) to determine the true range in X-ray emissions.

## 6. RESULTS AND DISCUSSION

As shown here, there is still much to learn about the most important class of pulsating variable stars – the Classical Cepheids. In terms of stellar evolution, our studies, as well as others, provide strong evidence of "real time" evolutionary changes taking place. This is particularly striking from the well-documented behavior of the nearest Cepheid - Polaris (see Neilson et al. 2012). As discussed previously, Polaris has been undergoing significant changes in its light curve, pulsation period and possibly in its luminosity. But this is not an isolated case, many other Cepheids, especially the more luminous, long period (higher mass) stars that have sufficiently long time baselines (> 50 years) show changes in their pulsation periods (Turner & Berdnikov 2004). This makes a strong case for continuing photometry and spectroscopy of the brighter galactic Cepheids that have observations going back several decades to search for subtle (and sometimes not so subtle –as in the case of Hubble Variable V19 in M33) changes arising from possible evolution effects. As an important part of the *SLiC* program, we are continuing photometry of over a dozen Cepheids.

In addition to the well-known optical variability of Cepheids, they are also found to have variable pulsation-phased UV line emissions. These emissions originate in plasmas with temperatures well above those of the Cepheid photospheres, showing Cepheids that possess extended atmospheres of heated, ionized plasma. The onset of enhanced line emissions (at phase ~0.8$\varphi$) is coincident with the beginning of the "piston phase" of each Cepheid – the phase at which the photosphere begins expanding from minimum radius. This leads us to favor the emergence and propagation of a shock, from the photosphere and through the surrounding atmosphere, as the most probable mechanism for the observed heating. This is in agreement with the study of Fokin et al. (1996), where spectroscopic observations of δ Cep were used to construct a pulsational model of this prototype Cepheid. The model shows shocks to pass through the surface of the Cepheid as it pulsates, the strongest of which passes through the surface right around $\varphi \approx 0.8$, the beginning of the enhanced UV line emissions. However, the Cepheids clearly display "quiescent" atmospheres at all observed phases. Because of this, we feel that additional smaller shocks could maintain a lower level of atmospheric heating throughout the all pulsational phases. However, a more likely explanation is that magnetic heating could be responsible for maintaining the "quiescent atmosphere" and the passage of a strong shock is then responsible for the phases of enhanced emission. Depending on the line species under study, the Cepheid emissions can vary by as much as roughly a factor of 10.

Our X-ray observations of Polaris, β Dor and δ Cep, show that Cepheids are indeed variable X-ray sources with $L_X \approx 6–12 \times 10^{28}$ ergs/s and $kT \approx 0.6–0.8$. However, originally there were doubts that the detected X-rays were not from the Polaris (and the other Cepheids) but could originate from coronal X-ray emissions from young cool star companions. As discussed previously, the multiple *HST-COS* FUV observations of δ Cep and β Dor carried out by us over the pulsation cycles of each star show a pronounced 7–10 fold peak in the FUV line emissions (particularly, in the N V 1240Å emission line which forms in hot $\sim 3 \times 10^5$ plasmas) near the expected ~0.8–1.0$\varphi$. The apparent variations of high temperature FUV emissions (such as N V) with pulsation phase, as well as evidence for the X-ray variability of Polaris and δ Cep, show that (in these cases) the Cepheids themselves are likely the X-ray sources. It will be important to secure X-ray observations covering the critical pulsation phases of 0.7–1.0$\varphi$, presumably when the X-ray activity should reach maximum strength.

In a broader context, recent multi-phase studies of Galactic and Magellanic Cloud Cepheids (Ngeow & Kanbur 2006) find the crucial Period-Luminosity (P-L) relation to display markedly different, nonlinear behavior during $\varphi \approx 0.70–0.95$. As shown in Fig. 5, this corresponds to the phase at which the FUV line emissions are strongest in δ Cep and β Dor, and also coincides with observed rapid radial velocity changes. The enhanced FUV emissions (and energy deposition) produce additional photospheric heating, with possible impacts on the P-L Law. The proposed *HST/COS* observations should also lead to much better characterizations of the shock-heating behavior and atmospheric outflow velocities in Cepheids, placing important constraints on Cepheid mass loss, a very important topic in terms of both the so-called "Cepheid Mass Discrepancy" and Cepheid evolution in general.

**FUTURE PROSPECTS**

In addition to our continuing ground-based photometry program, we have recently submitted an observing proposal to the *Hubble Space Telescope (HST)* for additional *HST/COS* FUV spectrophotometry of δ Cep and β Dor to better define the exact phase of maximum line emission. The observations are requested to cover the critical ~0.8–1.0 pulsation phases. Also, to expand the parameter space of observed Cepheids, we have requested HST/COS phase-sampled FUV spectrophotometry of the long period (P= 35.5-d), luminous, high-mass Cepheid, ℓ Car. To complement these FUV programs, we also have proposed for additional *Chandra* X-ray observations (and, in the future, *XMM-Newton*) of the Cepheid prototype/namesake, δ Cep, and the ~10-day "bump" Cepheid, β Dor, to expand phase coverage. When full X-ray and UV phase coverage is realized, we will carry out detailed models to characterize the shock behavior of the Cepheids, and emission measure analyses to detail their atmospheric plasmas (e.g. density, temperature distribution).

**ACKNOWLEDGEMENTS**

This paper was originally presented during August 2011 at the *Stars, Companions and their Interactions – A Memorial to Robert H. Koch* Conference. Although Bob is best known for his extensive work on close binary stars, as shown in papers presented at this conference, his interests and expertise extended to many other fields of Astronomy, including the Classical Cepheids discussed here. Both authors had frequent, useful discussions with Bob over the last several years in the evaluating the early photometry of Cepheids. In particular, Bob helped with the study and evaluation of historic magnitude measures of Polaris going back to the Almagest.   E.G. wishes to thank Bob for his friendship and for his thoughtful (and often frank) advice over 40 years since my time at the University of Pennsylvania as one of his first PhD graduate students. I looked forward to talking with Bob during his weekly visits to Villanova University during the latter part of his career. I enjoyed visiting Bob and Joanne and sometimes staying for their *nightly ice cream ritual*. Bob is deeply missed.

We gratefully acknowledge support from NASA grants this program: Spitzer grant JPL PID 40968, XMM Grant NNX08AX37G, HST Grants HST-GO-12302 and HST-GO-11726.

**REFERENCES**

Arellano Ferro, A. 1983, ApJ, 274, 755
Benedict, G. F., McArthur, B. E., Feast, M. W., Barnes, T. G., Harrison, T. E., Patterson, R. J., Menzies, J. W., Bean, J. L. & Freedman, W. L. 2007, AJ, 133, 1810
Böhm-Vitense, E. & Love, S. G. 1994, ApJ, 420, 401
Böhm-Vitense, E. 1958, ZA, 46, 108
Chabrier, G & Küker, M. 2006, A&A, 446, 1027
Cox, J. P. 1985, Cepheids: Theory and observations; Proceedings of the Colloquium, p. 126-146
Cox, J. P. 1974, Reports on Progress in Physics, 37, 563
DeWarf, L. E., Datin, K. M. & Guinan, E. F. 2010, ApJ, 722, 343
Eddington, A. S., Sir 1941, MNRAS, 101, 182
Engle, S. G., Guinan, E., Macri, L. & Pellerin, A. 2011, BAAS, 43, 2011
Engle, S. G., Guinan, E. F., Depasquale, J. & Evans, N. 2009, AIPC, 1135, 192
Engle, S. G., Guinan, E. F. & Koch, R. H. 2004, BAAS, 37, 378
Evans, N. R, Schaefer, G., Bond, H. E., Nelan, E., Bono, G., Karovska, M., Wolk, S., Sasselov, D., Guinan, E., Engle, S., Schlegel, E. & Mason, B. 2007, IAUS, 240, 102
Fokin, A.B. Gillet, D. & Breitfellner, M. G. 1996, A&A, 307, 503
Freedman, W. L. & Madore, B.F. 2010, ApJ, 719, 335

Kamper, K. W., & Fernie, J. D. 1998, AJ, 116, 936

Kraft, R. P 1957, ApJ, 125, 336

Kukarkin, B. V. 1975, in IPST Astrophysics Library (Keter Publishing House Jerusalem Ltd.)

Leavitt, Henrietta S. 1908, AnHar, 60, 87

Linsky, J. L., Wood, B. E., Judge, P., Brown, A., Andrulis, C. & Ayres, T. R. 1995, ApJ, 442, 381

Macri, L. M., Sasselov, D. D. & Stanek, K. Z. 2001, ApJ, 559, 243

Matthews, L. D., Marengo, M., Evans, N. R. & Bono, G. 2012, ApJ, 744, 53

Neilson, H. R., Engle, S. G., Guinan, E., Langer, N., Wasatonic, R. P. & Williams, D. B. 2012, ApJ, 745, 32

Ngeow, C.-C. & Kanbur, S. M. (2006), MNRAS, 369, 723

Redfield, S., Linsky, J. L., Ake, T. B., Ayres, T. R., Dupree, A. K., Robinson, R. D., Wood, B. E. & Young, Peter R. 2002, ApJ, 581, 626

Sasselov, D. D. & Lester, J. B. 1994, ApJ, 423, 795

Schmidt, E. G. & Parsons, S. B. 1982, ApJS, 48, 185

Schmidt, E. G. & Parsons, S. B. 1984a, ApJ, 279, 202

Schmidt, E. G. & Parsons, S. B. 1984b, ApJ, 279, 215

Spreckley, S. A. & Stevens, I. R. 2008, MNRAS, 388, 1239

Turner, D. G., Majaess, D. J., Lane, D. J., Balam, D. D., Gieren, W. P., Storm, J., Forbes, D. W., Havlen, R. J. & Alessi, B. 2012, MNRAS.tmp.2750T

Turner, D. G., Savoy, J., Derrah, J., Abdel-Sabour Abdel-Latif, M. & Berdnikov, L. N. 2005, PASP, 117, 207

Turner, D. G. & Berdnikov, L. N. 2004, A&A, 423, 335

van Leeuwen, F. 2007, A&A, 474, 653